# Time-dependent exchange-correlation current density functionals with memory

Yair Kurzweil and Roi Baer♦

*Department of Physical Chemistry and the Lise Meitner Minerva-Center for Computational Quantum Chemistry, the Hebrew University of Jerusalem, Jerusalem 91904 Israel.*

Most present applications of time-dependent density functional theory use adiabatic functionals, i.e. the effective potential at time *t* is determined solely by the density at the same time. This paper discusses a method that aims to go beyond this approximation, by incorporating "memory" effects: the potential will depend not only on present behavior but also on the past. In order to ensure the derived potentials are causal, we formulate the action on the Keldysh contour for electrons in electromagnetic fields, from which we derive suitable Kohn-Sham equations. The exchange correlation action is now a functional of the electron density and velocity field. A specific action functional is constructed which is Galilean invariant and yields a causal vector potential term to the Kohn-Sham equations that incorporates causal memory effects. We show explicitly that the exchange-correlation Lorentz force is zero. The potential is consistent with known dynamical properties of the homogeneous electron gas (in the linear response limit).

## I. INTRODUCTION

One of the important open theoretical challenges in molecular physics is the description of the dynamics of interacting electrons, for example, in a molecule or a cluster. In many typical cases the electrons start from their ground-state and are subject to an external perturbation, such as a longitudal electric field $\mathbf{E}_{ext}(\mathbf{r},t) = -\nabla\phi_{ext}(\mathbf{r},t)$ where $\phi_{ext}$ is an applied scala potential. Such a potential is imposed by the nuclei of the molecule or by external sources. Runge and Gross[1] proved that the time-dependent electron density $n(\mathbf{r},t)$ determines the external potential up to a purely time-dependent additive function. Thus the density uniquely determines the external electric field $\mathbf{E}_{ext}(\mathbf{r},t)$. Ghosh and Dhara[2] extended this result proving that the particle current-density $\mathbf{j}(\mathbf{r},t)$ determines the pair of external potentials $\phi_{ext}$ and $\mathbf{A}_{ext}$ up to an arbitrary gauge, i.e. the external electromagnetic fields $\mathbf{B}_{ext} = \nabla\times\mathbf{A}_{ext}$ $\mathbf{E}_{ext} = -\nabla\phi_{ext} - c^{-1}\dot{\mathbf{A}}_{ext}$ (where $c$ is the speed of light) are uniquely determined by the particle current density $\mathbf{j}$. This forms a basis for a class of successful approaches to electron dynamics, known collectively as time-dependent density functional theory (TDDFT)[1] and time-dependent current-density functional theory (TDCDFT) [2-4] which is analogous in many aspects to density functional theory (DFT) of Kohn and co-workers [5, 6].

In TDCDFT, the unique correspondence between the external fields and the particle current density $\mathbf{j}(\mathbf{r},t)$ is used to set up a map between interacting and non-interacting electron systems. Knowledge of $\mathbf{j}(\mathbf{r},t)$ in the interacting electron system is utilized to invent a system of non-interacting electrons that starting from their ground-state after being subject to an effective electromagnetic field $\mathbf{E}^{s}_{ext}$ and $\mathbf{B}^{s}_{ext}$ reconstruct the same evolving $\mathbf{j}(\mathbf{r},t)$. The class of current densities that can be generated in both an interacting and a non-interacting electron system is called "non-interacting v-representable current densities". For this class, the effective electromagnetic fields are unique[1, 2, 7] and therefore constitute a mapping of the interacting electron system upon a more tractable non-interacting one. In DFT, the issue of v-representability can be settled in a satisfactory manner, based on the quantum mechanical minimum principles, for a wide class of cases (see a discussion of this topic in [8]). An analogous result for TDDFT has not been established.

One way to formulate such a mapping, assuming it exists, for a given system is via an action principle. Because of the enormous complexity of the problem, only approximate mappings can be constructed in actual applications. The basic idea was outlined by Runge and Gross[1] (RG), who used the time-dependent quantum-mechanical action principle. The simplest approximations are "adiabatic" in the sense that they give a prescription to build $\mathbf{E}^{s}_{ext}$ and $\mathbf{B}^{s}_{ext}$ at time $t$ based solely on the current density at that same time. Such an approach is expected to be appropriate for slowly varying external fields. In general however, the effective fields at time $t$ depend on the current density also at *earlier times*, i.e. they must "have memory". Even in the linear response regime this is important, as excited states of double electron character cannot be accounted for using merely adiabatic functionals[9]. Memory effects are probably even more important for stronger fast varying fields, for example, the electronic correlation energy can be positive, an effect that cannot be accounted for using adiabatic functionals[10].

It is the purpose of this paper to present a simple prescription for constructing a Kohn-Sham mapping for TDCDFT which includes memory effects. To date, such a prescription is only partially available[3, 4]. A recently published method by Tokatly et al[11] also attempts to achieve this goal in a different way, based on the Landau Fermi-liquid theory where the local Lorentz force is a divergence of a stress tensor. In this paper we expand and give full detail the method which was recently proposed[12] allowing construction of a functional having the following properties:

♦ Corresponding author: FAX: +972-2-6513742, roi.baer@ huji.ac.il

a) **It is based on a TDCDFT action principle**. It is therefore robust and, one hopes, applicable beyond linear response. The present formulations of memory prescriptions[3, 4, 13] apply directly to potentials and usually cannot be derived from a 3-dimensional action principle.

b) **It is causal**, i.e. the external effective fields at time $t$ depend only on the *past* behavior of the system. Here we formulate the RG theory on a Keldysh contour. The resulting theory is thus causal. While resulting in a different functional, our approach is heavily based on the ideas and results of van Leeuwen[14].

c) **It obeys basic symmetry rules of quantum mechanics**, i.e. it does not allow electrons to exert a net force upon themselves and it is gauge invariant. The first of these conditions, termed *Galilean invariance* [15, 16]. Galilean invariance ensures compliance of TDDFT dynamics with the Harmonic potential theorem of Dobson[17].

d) **It is consistent with known static and dynamical properties of the homogenous electron gas** (HEG) when subject to weak external electromagnetic fields. Here we follow parameterizations of such properties [18-20].

The idea of assembling an approach that is Galilean invariant (GI) and is consistent with the linear response properties of the HEG goes back to the work of Dobson Brunner and Gross[3]. Our approach however is different in several aspects, most notably, we give here a fully 3D prescription, treating both the transverse as well as longitudal response properties.

The mathematical intricacies of such a functional are rather involved. We find, that a relatively transparent, tractable and formally appealing formulation is obtained when the functionals are made to depend on the velocity field $\mathbf{u} = \mathbf{j}/n$ and density $n$ instead of on the current density $\mathbf{j}$. The utility of using the velocity field has been stressed by Vingale et al [4, 13], who use it to build a linear response TDCDFT approach. Some applications showing the utility of the memory effects have been recently published[21, 22].

With the velocity field $\mathbf{u}(\mathbf{r},t)$, the concept of "fluid parcel trajectory" is naturally defined[3, 4, 17]. Thus the electronic density is viewed as a fluid, and each fluid-parcel flows along a trajectory. The memory effects are easily described within a parcel, because there they are "local". This is a Lagrangian description of the electron dynamics and should be contrasted with the "Eularian" system, where a fixed coordinate system is used to describe the density. Finally, since other parts of the theory are more naturally described in the Eularian system, a Lagrangian-to-Eularian transformation must be made after formulation of the relevant exchange-correlation potentials.

The structure of the paper is as follows. In section II we formulate a causal theory of TDCDFT, based on a similar approach for TDDFT[14]. Next we introduce a specific action functional including memory effects (section III). In section IV we take the appropriate functional derivatives to obtain the exchange-correlation potentials. The parameterization of the functionals kernel functions is derived in section V. A discussion and summary is then given in sectionVI.

## II. TDCDFT ACTION

In TDDFT time-dependent Schrödinger equations are set up for one-particle orbitals from which the density and current density can be constructed. This is similar to DFT, where time-independent equations are solved for non-interacting electrons in an effective potential composed of the external potential, the Hartree potential and the exchange correlation potential [6]. In DFT the exchange-correlation (XC) energy functional $E_{xc}[n_0]$ ($n_0(\mathbf{r})$ being the ground-state number-density of the electronic system) is a basic concept, determining the exchange-correlation potential through functional derivation, $v_{xc}(\mathbf{r}) = \delta E_{xc}[n_0]/\delta n_0(\mathbf{r})$. In TDDFT the analogous concept, introduced by Runge and Gross (RG) [1], is the XC action functional $S_{xc}[n]$, where $n(\mathbf{r},t)$ is the TD number-density. RG assumed that functional derivation of $S_{xc}$ yields the exchange correlation potential as in DFT. One very successful case is TD adiabatic local density approximation (TD-ALDA), where the DFT local density approximation (LDA) functional is used as a memory-less action. TD-ALDA is very successful for computing some dynamical properties of molecules[23-27]. Yet, it is inadequate in processes that involve motion of electrons over long distances [22, 28] or in cases of double excitations [29]. Thus, it is important to develop methods beyond the adiabatic assumption and several attempts in this direction have been made [4, 10, 30, 31].

The RG procedure, as well as any attempt to derive the potentials from a time dependent functional of the density was found to violate causality [14]. This can be remedied by formulating the action on a Keldysh contour, which was used earlier in the context of TDDFT[14, 32, 33]. The work of van Leeuwen[14] has especially elegant way of dealing with causality using the Keldysh contour. We now generalize this method for the magnetic field case, obtaining a formulation of TDCDFT. Consider a system of $N_e$ electrons under an external electromagnetic field. We specifically focus on the dynamics in a time interval $[0,t_f]$. The Keldysh contour is a parameterization $t(\tau)$, using a parameter $\tau \in [0,\tau_f]$. This function $t(\tau)$ taken from $0$ to $t_f$ and then back to zero: $t(0) = t(\tau_f) = 0$. We start from the Schrödinger equation of a system of $N_e$ electrons on a Keldysh contour:

$$i\dot{t}^{-1}\partial_\tau \Psi(\tau) = \hat{H}(\tau)\Psi(\tau) \qquad \Psi(0) = \Psi_0 \qquad (2.1)$$

The Hamiltonian here is of the general form, allowing external electromagnetic fields:

$$\hat{H}(\tau) = \sum_{i=1}^{N_e}\left[\frac{(\mathbf{p}_i + \mathbf{a}(\mathbf{r}_i,t))^2}{2} + v_{ext}(\mathbf{r}_i,t)\right] + \frac{1}{2}\sum_{i\neq j}\frac{1}{r_{ij}} \quad (2.2)$$

Despite the formal similarity to the Schrödinger equation, Eq. (2.1) is different since it is solved on the contour. To recover the physical equation, we must introduce physical potentials on the contour, i.e. $v(\mathbf{r},t) = v_p(\mathbf{r},\tau(t))$ and $\mathbf{a}(\mathbf{r},t) = \mathbf{a}_p(\mathbf{r},\tau(t))$. Only then will the solution be of the form $\Psi(t) = \Psi_p(\tau(t))$ where $\Psi_p(\tau)$ is the physical wave-function.

We introduce the following action:

$$A[\tilde{\mathbf{a}}] = \mathrm{Re}\, i\ln\langle\Psi_0|U(t_f,0)|\Psi_0\rangle \quad (2.3)$$

Where the evolution operator is a solution of:

$$i\dot{t}^{-1}\partial_t U(t,0) = \left[\sum_{i=1}^{Ne}\frac{(\mathbf{p}_i + \tilde{\mathbf{a}}(\mathbf{r}_i))^2}{2} + \frac{1}{2}\sum_{i\neq j}\frac{1}{r_{ij}}\right]U(t,0) \quad (2.4)$$

With $U(0,0)$ the identity operator and:

$$\tilde{\mathbf{a}}(\mathbf{r},t) = \mathbf{a}(\mathbf{r},t) - \int_0^t \nabla v_{ext}(\mathbf{r},t')\,dt' \quad (2.5)$$

is the vector potential in the fixed gauge we are going to use. It is easy to show that in this gauge:

$$\mathbf{j}(\mathbf{r},t) \equiv \frac{\delta A}{\delta\tilde{\mathbf{a}}(\mathbf{r},t)} = \mathrm{Re}\frac{\langle\Psi_0|U(t_f,0)\hat{\mathbf{j}}_H(\mathbf{r},t)|\Psi_0\rangle}{\langle\Psi_0|U(t_f,0)|\Psi_0\rangle} \quad (2.6)$$

With:

$$\hat{\mathbf{j}}_H(\mathbf{r},t) = U(0,t)\,\hat{\mathbf{j}}\,U(t,0) \quad (2.7)$$

and:

$$\hat{\mathbf{j}}(\mathbf{r}) = \sum_{i=1}^{N_e}(\hat{\mathbf{p}}_i + \tilde{\mathbf{a}}(\hat{\mathbf{r}}_i))\delta(\mathbf{r}-\hat{\mathbf{r}}_i) + c.c \quad (2.8)$$

the particle current density operator is . One can also define the "density":

$$n(\mathbf{r},t) \equiv \mathrm{Re}\frac{\langle\Psi_0|U(t_f,0)\hat{n}_H(\mathbf{r},t)|\Psi_0\rangle}{\langle\Psi_0|U(t_f,0)|\Psi_0\rangle} \quad (2.9)$$

It is straightforward to show that the continuity equation holds:

$$\dot{t}^{-1}\partial_t n + \nabla\cdot\mathbf{j} = 0 \quad (2.10)$$

Furthermore, when a physical vector potential is plugged in the expressions, $\mathbf{j}$ and $n$ become the corresponding gauge-invariant physical functions.

One can now perform a Legendre transform:

$$\tilde{S}[\mathbf{j}] = A[\tilde{\mathbf{a}}] - \int d^3r\int_C dt\,\mathbf{j}(\mathbf{r},t)\cdot\tilde{\mathbf{a}}(\mathbf{r},t) \quad (2.11)$$

Functional differentiation gives:

$$\frac{\delta\tilde{S}}{\delta\mathbf{j}(\mathbf{r},t)} = -\tilde{\mathbf{a}}(\mathbf{r},t) \quad (2.12)$$

We then write this functional in the KS way, defining the exchange correlation action $\tilde{S}_{XC}$:

$$\tilde{S}[\mathbf{j}] = S_0[\mathbf{j}] + S_H[n] + \tilde{S}_{XC}[\mathbf{j}]. \quad (2.13)$$

$S_0$ is the action for a set of non-interacting of electrons and:

$$S_H[n] = \tfrac{1}{2}\iint d^3r\,d^3r'\int_C dt\,\frac{n(\mathbf{r},t)\,n(\mathbf{r}',t)}{|\mathbf{r}-\mathbf{r}'|}, \quad (2.14)$$

The functional $\tilde{S}$ in Eq. (2.13) will be awkward to handle because of the functional dependence of $n$ on $\mathbf{j}$ through the continuity equation (2.10). A better approach would be to define a functional of $n$ and $\mathbf{j}$, both treated as independent variables, and add (2.10) as a constraint. Furthermore, we will need in the next sections a functional expressed in terms of $n$ and the velocity field $\mathbf{u} = \mathbf{j}/n$, thus we define $S[n,\mathbf{u}]$:

$$S[n,\mathbf{u}] = \tilde{S}[n\mathbf{u}] + \int w(\mathbf{r},t)\left[\dot{t}^{-1}\partial_t n(\mathbf{r},t) + \nabla\cdot(n(\mathbf{r},t)\mathbf{u}(\mathbf{r},t))\right] \quad (2.15)$$

The functional derivatives, for $0 < t < t_f$ are:

$$\frac{\delta S}{\delta\mathbf{u}(\mathbf{r},t)} = -(\tilde{\mathbf{a}}(\mathbf{r},t) + \nabla w(\mathbf{r},t))\,n(\mathbf{r},t) \quad (2.16)$$

And:

$$\frac{\delta S}{\delta n(\mathbf{r},t)} = -(\tilde{\mathbf{a}}(\mathbf{r},t) + \nabla w(\mathbf{r},t))\cdot\mathbf{u}(\mathbf{r},t) - \dot{t}^{-1}\partial_t w(\mathbf{r},t) \quad (2.17)$$

At $t = 0, t_f$ there are additional delta-function terms. These disappear when physical densities are used, thus we do not consider them.

We may analogously define the exchange correlation action $S_{XC}[n,\mathbf{u}]$ using Eqs. (2.13) and (2.15) by:

$$S[n,\mathbf{u}] = S_0[n\mathbf{u}] + S_H[n] + S_{XC}[n,\mathbf{u}] \quad (2.18)$$

The non-interacting action is $S_0[n,\mathbf{u}]$ in this case is tractable, since we may let the electrons evolve under the potential $\tilde{\mathbf{a}}_s$ according to the Schrodinger equations $\phi_k$:

$$i\dot{t}^{-1}\partial_t\phi_k(\mathbf{r},t) = \frac{1}{2}(\hat{\mathbf{p}} + \tilde{\mathbf{a}}_s(\mathbf{r},t))^2\phi_k(\mathbf{r},t). \quad (2.19)$$

From Eq. (2.12), applied to the non-interacting electrons:

$$\frac{\delta S_0}{\delta \mathbf{u}(\mathbf{r},t)} = -\left(\tilde{\mathbf{a}}_s(\mathbf{r},t) + \nabla w_s(\mathbf{r},t)\right) n(\mathbf{r},t)$$
$$\frac{\delta S_0}{\delta n(\mathbf{r},t)} = -\left(\tilde{\mathbf{a}}_s(\mathbf{r},t) + \nabla w_s(\mathbf{r},t)\right)\cdot \mathbf{u}(\mathbf{r},t) - \dot{\tau}^{-1}\partial_t w_s(\mathbf{r},t) \quad (2.20)$$

Equating the functional derivatives of (2.18) leads to

$$-n\left[(\tilde{\mathbf{a}} - \tilde{\mathbf{a}}_s) + \nabla(w - w_s)\right] = \tilde{\mathbf{a}}_{XC}$$
$$-\mathbf{u}\cdot\left[(\tilde{\mathbf{a}} - \tilde{\mathbf{a}}_s) + \nabla(w - w_s)\right] - \partial_t(w - w_s) = v_H + v_{XC} \quad (2.21)$$

With $w_s$ the Lagrange multipliers for non-interacting electrons, and $\mathbf{a}_{XC} = \delta S_{XC}/\delta\mathbf{u}$ and $v_{XC} = \delta S/\delta n$. This leads to:

$$\tilde{\mathbf{a}}_s = \tilde{\mathbf{a}} + \tilde{\mathbf{a}}_{XC}/n + \nabla W$$
$$W = \int_0^t \left[\mathbf{u}\cdot\tilde{\mathbf{a}}_{XC}/n - v_H - v_{XC}\right]\dot{\tau}^{-1}(t')\,dt' \quad (2.22)$$

Where $W = w - w_s$. The formalism is gauge invariant. One can apply any gauge transform to Eq. (2.19) and transform part of the longitudinal component of the vector potential into an external potential. This corresponds to a different choice of $W$ in Eq. (2.22). Indeed, for constructing definite Kohn Sham equations, a certain gauge must usually be chosen. Once this is done, the equations have the form:

$$i\partial_t \varphi_k(\mathbf{r},t) = \left[\frac{1}{2}\left(\hat{\mathbf{p}} + \mathbf{a}_s(\mathbf{r},t)\right)^2 + v_s(\mathbf{r},t)\right]\varphi_k(\mathbf{r},t). \quad (2.23)$$

Along with:

$$\mathbf{a}_s = \mathbf{a} + \mathbf{a}_{XC}/n$$
$$v_s = v_{ext} + v_H + v_{XC} - \mathbf{u}\cdot\mathbf{a}_{XC}/n \quad (2.24)$$

Eqs. (2.23) and (2.24) form a set of equations which must be solved self-consistently. The basic issue now is the construction of an approximation to the exchange correlation action $S_{XC}[n,\mathbf{u}]$, which we do next.

## III. ACTION WITH MEMORY

We now describe the basic principles of our approach for building an action principle for TDCDFT with memory. For simplicity, we assume the interacting system is not subject to a magnetic field, thus $\mathbf{a} = 0$. We separate the functional to an instantaneous response part and a memory part:

$$S_{XC} = S_A + S_{GIXC} \quad (3.1)$$

We assume $S_A$ is the ALDA , i.e. the functional that yields the following xc-potential:

$$v_{xc,ALDA}(\mathbf{r},t) = \frac{d}{dn}\left(\varepsilon_{LDA}(n)n\right)\Big|_{n=n(\mathbf{r},t)} \quad (3.2)$$

Where $\varepsilon_{xc}$ is the exchange-correlation energy per particle in the homogeneous electron gas (HEG) at its ground-state.

The functional $S_{GIXC}$ will be expressed in terms of quantities that are zero at time zero, before any external time-dependent perturbation is applied. It is a functional not only of the density, but also the electron fluid velocity-field $\mathbf{u}(\mathbf{r},t) = \mathbf{j}(\mathbf{r},t)/n(\mathbf{r},t)$, where $\mathbf{j}(\mathbf{r},t)$ is the current density of the fluid[13]. The velocity is initially zero (we assume no external magnetic fields at time zero). In a more general treatment, $S_A$ will also include the stationary velocity field which exists in a static magnetic field, but here we assume it is a functional of the density only.

The functional must observe causality. Van-Leeuwen proved[14] that an action principle of the form discussed in the previous section violates causality. Thus a change in the formulation is required. This is done, by introducing a pseudo-time $\tau$ and a function $t(\tau)$, which starts at $t = 0$ changes to the final time $t_f$ and then is driven backwards – from $t_f$ to 0. The way we use the pseudo-time Keldysh technique is discussed in Appendix A. Based on that technique we set out to construct a functional $S_{GIXC}[n,\mathbf{u}]$ from which the xc-potentials can be inferred by functional derivation (the actual derivatives and potentials are given in section IV):

$$v_{GIXC}(\mathbf{r},t) = \frac{\delta S_{GIXC}[n,\mathbf{u}]}{\delta n(\mathbf{r},\tau)}\Big|_p,$$
$$\mathbf{a}_{GIXC}(\mathbf{r},t) = \frac{\delta S_{GIXC}[n,\mathbf{u}]}{\delta \mathbf{u}(\mathbf{r},\tau)}\Big|_p. \quad (3.3)$$

Where the subscript $p$ denotes evaluation at the physical density $n_p$ and velocity-field $\mathbf{u}_p$. A function $f(\tau)$ is called "physical" if there exists a function $f_p(t)$ such that $f(\tau) = f_p(t(\tau))$, see Appendix A for more details. These potentials are used in the Kohn Sham scheme, (2.23) where the effective potentials are, analogous to Eq. (2.24) (with $\mathbf{a} = 0$):

$$v_s(\mathbf{r},t) = v_p(\mathbf{r},t) + v_H(\mathbf{r},t) + v_{ALDA}(\mathbf{r},t) + v_{GIXC}(\mathbf{r},t) - \mathbf{u}(\mathbf{r},t)\cdot\mathbf{a}_{GIXC}(\mathbf{r},t)/n(\mathbf{r},t) \quad (3.4)$$
$$\mathbf{a}_s(\mathbf{r},t) = \mathbf{a}_{GIXC}(\mathbf{r},t)/n(\mathbf{r},t)$$

The potentials $v_{GIXC}$ and $\mathbf{a}_{GIXC}$ should be derived from a GI action functional, the definition of which we discuss now. Consider two coordinate systems, $\mathbf{r}$ of the "lab" frame and $\tilde{\mathbf{r}}$ of a "moving" frame, where

$$\tilde{\mathbf{r}} = \mathbf{r} - \mathbf{x}(\tau), \quad (3.5)$$

and $\mathbf{x}(\tau)$ is an arbitrary (accelerated) trajectory (with $\mathbf{x}(0) = 0$ for simplicity) in pseudo-time accelerated frame. A functional $S[n,\mathbf{u}]$ is considered GI if :

$$S[\tilde{n},\tilde{\mathbf{u}}] = S[n,\mathbf{u}] \quad (3.6)$$

where $\tilde{n}(\mathbf{r},t)$ and $\tilde{\mathbf{u}}(\mathbf{r},t)$ are the density and velocity-field in the accelerated frame:

$$\begin{aligned}\tilde{n}(\mathbf{r},t) &= n(\mathbf{r}+\mathbf{x}(t),t),\\ \tilde{\mathbf{u}}(\mathbf{r},t) &= \mathbf{u}(\mathbf{r}+\mathbf{x}(t),t) - \dot{\tau}(t)^{-1}\dot{\mathbf{x}}(t).\end{aligned} \quad (3.7)$$

When a physical velocity-field $\mathbf{u}(\mathbf{r},t)=\mathbf{u}_p(\mathbf{r},\tau(t))$ and trajectory $\mathbf{x}(t)=\mathbf{x}_p(\tau(t))$ are used in Eq. (3.7) the usual Galilean transformation is recovered:

$$\tilde{\mathbf{u}}_p(\mathbf{r},\tau) = \mathbf{u}_p(\mathbf{r}+\mathbf{x}(\tau),\tau) - \dot{\mathbf{x}}_p(\tau). \quad (3.8)$$

According to Newton's third law, electrons should not produce a net force upon themselves. This requirement imposes a strict condition on the XC potentials, as discussed in ref. [16], namely that the net XC force on the electrons is zero. Now, what are the XC forces in the present theory? Since the XC forces are of formal similarity to electromagnetic potentials, the force they produce per volume should be recovered from the Lorentz force of electromagnetic theory. The Lorentz force is discussed in detail in ref. [34], where it is also shown that electromagnetic fields are related to their potentials by: $\mathbf{E} = -\nabla\phi - c^{-1}\dot{\mathbf{A}}$ and $\mathbf{B} = \nabla\times\mathbf{A}$. In the case of a flowing charge distribution in an electromagnetic field, the Lorentz force per volume exerted on a charge distribution $\rho$ flowing with velocity $\mathbf{u}$ is:

$$\mathbf{F}_L = \rho\{\mathbf{E}+\mathbf{u}\times\mathbf{B}\} \quad (3.9)$$

In our case, the XC potentials are defined slightly differently, an analogy with electromagnetism is obtained by setting $c=1$, $\phi\to -v$ and $\mathbf{A}\to\mathbf{a}$ and $\rho\to n$. The form for the XC potentials given in Eq. (3.4) leads to effective fields given by:

$$\begin{aligned}\mathbf{E}_{GIXC} &= \nabla(v_{GIXC} - \tilde{\mathbf{a}}_{GIXC}\cdot\mathbf{u}) - \dot{\tilde{\mathbf{a}}}_{GIXC}\\ \mathbf{B}_{GIXC} &= \nabla\times\tilde{\mathbf{a}}_{GIXC}\end{aligned} \quad (3.10)$$

Where $n\tilde{\mathbf{a}}_{GIXC} = \mathbf{a}_{GIXC}$. Plugging the fields of Eq. (3.10) into Eq. (3.9), yields the XC force density. In Appendix B a TDCDFT generalization of the results of ref. [16] is given, showing Galilean invariance implies that the net XC force defined this way is zero.

In order to impose Galilean invariance on the XC action functional we use a Lagrangian coordinate system[17]. We introduce the trajectory function $\mathbf{R}(\mathbf{r},t)$, describing the position at pseudo-time $t$ of an electron fluid parcel, which at $t=0$ was at $\mathbf{r}$, obeying the equations of motion:

$$\frac{\partial\mathbf{R}(\mathbf{r},t)}{\partial t} = \dot{\tau}(t)\,\mathbf{u}(\mathbf{R}(\mathbf{r},t),t), \qquad \mathbf{R}(\mathbf{r},0)=\mathbf{r}, \quad (3.11)$$

where $\mathbf{u}(\mathbf{r},t)$ is the velocity field. When a physical velocity field is used, $\mathbf{u}(\mathbf{r},t)=\mathbf{u}_p(\mathbf{r},\tau(t))$, this definition is compatible with the physical trajectory $d\mathbf{R}_p/dt=\mathbf{u}_p$ where $\mathbf{R}(\mathbf{r},t)=\mathbf{R}_p(\mathbf{r},\tau(t))$. The Galilean transform of $\mathbf{R}$ is: $\tilde{\mathbf{R}}(\mathbf{r},t)=\mathbf{R}(\mathbf{r},t)-\mathbf{x}(t)$ (*note:* $\tilde{\mathbf{R}}(\tilde{\mathbf{r}},t)=\mathbf{R}(\mathbf{r},t)-\mathbf{x}(t)$, because $\mathbf{r}$ is the position at time $t=0$, when the two frames are identical). Following refs [3, 17], we introduce the Lagrangian velocity-field $\mathbf{U}(\mathbf{r},t)=\mathbf{u}(\mathbf{R}(\mathbf{r},t),t)$ and density $N(\mathbf{r},t)=n(\mathbf{R}(\mathbf{r},t),t)$, noting that:

$$\tilde{N}(\mathbf{r},t)=N(\mathbf{r},t);\qquad \tilde{\mathbf{U}}(\mathbf{r},t)=\mathbf{U}(\mathbf{r},t)-\dot{\mathbf{x}}(t). \quad (3.12)$$

Since $N=\tilde{N}$ and $\nabla\mathbf{U}=\nabla\tilde{\mathbf{U}}$, any functional of $N$ and $\nabla\mathbf{U}$ is trivially GI (i.e. obeys $S[N,\mathbf{U}]=S[\tilde{N},\tilde{\mathbf{U}}]$), we can write a general GIXC action as follows:

$$\begin{aligned}S_{GIXC}[n,\mathbf{u}] = \int d^3r\int_0^{t_f}\dot{\tau}(t')dt'\int_0^{t'}\dot{\tau}(t'')dt''\\ \Pi\left(N(\mathbf{r},t'),N(\mathbf{r},t''),\nabla\mathbf{U}(\mathbf{r}',t'),\nabla\mathbf{U}(\mathbf{r}',t''),t',t''\right),\end{aligned} \quad (3.13)$$

Where $\Pi\left(N',N'',\nabla\mathbf{U}',\nabla\mathbf{U}'',t',t''\right)$ is an appropriate kernel functional. This equation opens the door for a generalized 'gradient' approximation in the time domain. Furthermore, this general XC action is not limited to the linear perturbation regime. To demonstrate the practicality of this GIXC action, we choose a simple non-trivial functional, which in the linear response regime coincides with known forms[35]:

$$\begin{aligned}S_{GIXC}[n,\mathbf{u}] = \int d^3r'\int_0^{t_f}\dot{\tau}(t')dt'\int_0^{t'}\dot{\tau}(t'')dt''\\ \Big\{F_L\left(N(\mathbf{r}',t''),\tau(t')-\tau(t'')\right)\nabla\cdot\mathbf{U}(\mathbf{r}',t')\,\nabla\cdot\mathbf{U}(\mathbf{r}',t'')+\\ F_T\left(N(\mathbf{r}',t''),\tau(t')-\tau(t'')\right)\nabla\times\mathbf{U}(\mathbf{r}',t')\cdot\nabla\times\mathbf{U}(\mathbf{r}',t'')\Big\}\end{aligned}$$

(3.14)

The response kernels $F^L$ and $F^T$, analogous to the xc-energy per particle in LDA, carry the information of a generic physical system, such as that of the homogeneous electron gas (HEG). The functions $F^{L,T}$ can then be obtained from approximated form of the LR response functions, as discussed in the following section.

## IV. EXCHANGE CORRELATION POTENTIALS

When physical density and velocity field are plugged into Eq. (3.14), it can be shown that $S_{GIXC}$ is identically zero (use Eq. (A.4) for proving this). This does not mean that the potential derived from it are zero though because the functional derivative is done with respect to *any* (not necessarily physical density and velocity field).

In order to obtain the GIXC scalar and vector-potentials, $\mathbf{a}_{\mathrm{GIXC}}(\mathbf{r},t)=\delta S_{\mathrm{GIXC}}/\delta\mathbf{u}(\mathbf{r},t)$, $v_{\mathrm{GIXC}}(\mathbf{r},t)=\delta S_{\mathrm{GIXC}}/\delta n(\mathbf{r},t)$, we need to first to compute the position and velocity Jacobians. The trajectory-position Jacobian matrix is calculated in Appendix C, resulting in:

$$\Im_{ij}(\mathbf{r},t) = [\nabla \mathbf{R}]_{ij}(\mathbf{r},t) = \partial \boldsymbol{R}_i(\mathbf{r},t)/\partial \boldsymbol{r}_j . \qquad (4.1)$$

The trajectory-position Jacobian matrix $\Im(\mathbf{r},\boldsymbol{\tau})$ tells us how a path is affected when its initial position is perturbed. In actual computations it can be determined directly from Eq. (4.1). Furthermore, from Eqs. (4.1) and (3.11), $\dot{\Im}(\mathbf{r},t) = \dot{\boldsymbol{\tau}}(t)\nabla\mathbf{U}(\mathbf{r},t)$, from which $\dot{\Im}(\mathbf{r},t) = \dot{\boldsymbol{\tau}}(t)[\nabla\mathbf{u}](\mathbf{R}(\mathbf{r},t),t)\Im(\mathbf{r},t)$ (where $[\nabla\mathbf{u}]_{ij} = \partial \boldsymbol{u}_i(\mathbf{r},t)/\partial \boldsymbol{r}_j$), thus the Jacobian matrix is given by:

$$\Im(\mathbf{r},t) = \wp \exp\left[\int_0^t [\nabla\mathbf{u}]\left(\mathbf{R}(\mathbf{r},t'),t'\right)\dot{\boldsymbol{\tau}}(t')\boldsymbol{d}t'\right] \qquad (4.2)$$

where $\wp$ is the $t$-ordering symbol (earlier times to the right). The Jacobian determinant ensures particle conservation by correcting for particle density when physical function $\boldsymbol{N}(\mathbf{r},t) = \boldsymbol{N}_p(\mathbf{r},\boldsymbol{\tau}(t))$ are used (see Appendix C): $|\Im(\mathbf{r},\boldsymbol{\tau})| = \boldsymbol{N}_p(\mathbf{r},0)/\boldsymbol{N}_p(\mathbf{r},\boldsymbol{\tau})$.

Next, we need also the trajectory-velocity Jacobian matrix:

$$\mathbf{G}_{ij}(\mathbf{r}',t';\mathbf{r},t) = \delta\mathbf{R}_i(\mathbf{r}',t')/\delta\mathbf{u}_j(\mathbf{r},t) \qquad (4.3)$$

This function tells us how a trajectory originating at $\mathbf{r}'$ changes at time $t'$ as a result of a perturbation in the velocity field applied at position $\mathbf{r}$ at time $t$. Taking the derivative with respect to $t'$ of $\mathbf{G}_{ij}$, one arrives after some manipulations at:

$$\dot{\boldsymbol{\tau}}(t')^{-1}\frac{\partial}{\partial t'}\mathbf{G}_{ij} = \delta_{ij}\delta(t'-t)\delta\left(\mathbf{R}(\mathbf{r}',t')-\mathbf{r}\right) + [\nabla_k \mathbf{u}_i]\left(\mathbf{R}(\mathbf{r}',t'),t'\right)\mathbf{G}_{kj}$$

This equation of motion for $\mathbf{G}$ can be readily solved using Eq. (4.2):

$$\mathbf{G}_{ij}(\mathbf{r}',t';\mathbf{r},t) = \left[\Im(\mathbf{r}',t')\Im(\mathbf{r}',t)^{-1}\right]_{ij} \times \theta(t'-t)\delta\left(\mathbf{R}(\mathbf{r}',t)-\mathbf{r}\right) \qquad (4.4)$$

In this expression it is evident that the only trajectory affected by the perturbation is that which passes at the space-time point of the velocity-perturbation. $\theta(t'-t)$ is the Heaviside function enforcing causality: only the future trajectory is affected. The position Jacobians play the role of "propagators". $\Im(\mathbf{r}',t)^{-1}$ propagates backward from perturbation time $t$ to time zero and $\Im(\mathbf{r}',t')$ forward from time zero to "present" time $t'$. An important property of $\mathbf{G}$ is its spatial-sparsity: it is strictly zero unless $\mathbf{r}$ and $\mathbf{r}'$ refer to the *same fluid element.* $\mathbf{G}$ is non-local in time, but in any application the memory functionals require only a limited time non-locality. Thus $\mathbf{G}$ can be computed on the fly. With the Jacobians, we can write the functional derivatives, determining how changes in the Eularian fields affect the Lagrangian variables:

$$\delta\boldsymbol{N}(\mathbf{r}',t')/\delta\boldsymbol{n}(\mathbf{r},t) = \delta\left(\mathbf{R}(\mathbf{r}',t)-\mathbf{r}\right)\delta(t-t')/\dot{\boldsymbol{\tau}}$$

$$\delta\boldsymbol{N}(\mathbf{r}',t')/\delta\mathbf{u}(\mathbf{r},t) = \nabla\boldsymbol{N}(\mathbf{r}',t')\Im(\mathbf{r}',t')^{-1}\mathbf{G}(\mathbf{r}',t';\mathbf{r},t)$$

$$\delta\mathbf{U}(\mathbf{r}',t')/\delta\boldsymbol{n}(\mathbf{r},t) = 0,$$

$$\delta\mathbf{U}_i(\mathbf{r}',t')/\delta\mathbf{u}_j(\mathbf{r},t) = \dot{\boldsymbol{\tau}}(t')^{-1}\partial\mathbf{G}_{ij}(\mathbf{r}',t';\mathbf{r},t)/\partial t',$$

From these relations, we compute the potentials:

$$\frac{\delta \boldsymbol{S}_{GIXC}}{\delta\boldsymbol{n}(\mathbf{r},t)} = \int_0^{t_f}\boldsymbol{V}(\bar{\mathbf{r}},\bar{t})\frac{\delta\boldsymbol{N}(\bar{\mathbf{r}},\bar{t})}{\delta\boldsymbol{n}(\mathbf{r},t)}\boldsymbol{d}^3\bar{\boldsymbol{r}}\dot{\boldsymbol{\tau}}(\bar{t})\boldsymbol{d}\bar{t},$$

$$\frac{\delta \boldsymbol{S}_{GIXC}}{\delta\mathbf{u}(\mathbf{r},t)} = \int_0^{t_f}\left[\mathbf{A}(\bar{\mathbf{r}},\bar{t})\frac{\delta\mathbf{U}(\bar{\mathbf{r}},\bar{t})}{\delta\mathbf{u}(\mathbf{r},t)} + \boldsymbol{V}(\bar{\mathbf{r}},\bar{t})\frac{\delta\boldsymbol{N}(\bar{\mathbf{r}},\bar{t})}{\delta\mathbf{u}(\mathbf{r},t)}\right]\boldsymbol{d}^3\bar{\boldsymbol{r}}\dot{\boldsymbol{\tau}}(\bar{t})\boldsymbol{d}\bar{t}.$$

where $\boldsymbol{V}(\mathbf{r},t) = \delta\boldsymbol{S}_{GIXC}/\delta\boldsymbol{N}(\mathbf{r},t)$ and $\mathbf{A}(\mathbf{r},t) = \delta\boldsymbol{S}_{GIXC}/\delta\mathbf{U}(\mathbf{r},t)$. It is straightforward to verify, using Eqs. (A.3) and (A.5), that when physical densities and velocities are plugged into these expressions and the causality property $\boldsymbol{F}_{L,T}(\boldsymbol{N},t<0) = 0$ enforced, the vector potential comes out causal, having both transverse and longitudal parts:

$$\mathbf{a}_{GIXC}\left(\mathbf{R}(\mathbf{r},\boldsymbol{\tau}),\boldsymbol{\tau}\right) = |\Im(\mathbf{r},\boldsymbol{\tau})|^{-1}\Im(\mathbf{r},\boldsymbol{\tau})^{-1}\times \int_0^{\boldsymbol{\tau}}\boldsymbol{d}\boldsymbol{\tau}'\dot{\mathbf{A}}_p(\mathbf{r},\boldsymbol{\tau}')\Im(\mathbf{r},\boldsymbol{\tau}'), \qquad (4.5)$$

where

$$\mathbf{A}_p(\mathbf{r},\boldsymbol{\tau}) = -\nabla\int_0^{\boldsymbol{\tau}}\boldsymbol{d\tau}'\boldsymbol{F}_L\left(\boldsymbol{N}(\mathbf{r},\boldsymbol{\tau}'),\boldsymbol{\tau}-\boldsymbol{\tau}'\right)\nabla\cdot\mathbf{U}(\mathbf{r},\boldsymbol{\tau}') + \nabla\times\int_0^{\boldsymbol{\tau}}\boldsymbol{d\tau}'\boldsymbol{F}_T\left(\boldsymbol{N}(\mathbf{r},\boldsymbol{\tau}'),\boldsymbol{\tau}-\boldsymbol{\tau}'\right)\nabla\times\mathbf{U}(\mathbf{r},\boldsymbol{\tau}'), \qquad (4.6)$$

Furthermore, the GI XC-potential is identically zero:

$$\boldsymbol{v}_{GIXC}\left(\mathbf{R}(\mathbf{r},\boldsymbol{\tau}),\boldsymbol{\tau}\right) = |\Im(\mathbf{r},\boldsymbol{\tau})|^{-1}\boldsymbol{V}_p(\mathbf{r},\boldsymbol{\tau}) = 0 \qquad (4.7)$$

Notice that the left hand side of Eq. (4.5) gives the vector potential at $\mathbf{a}_{GIXC}\left(\mathbf{R}(\mathbf{r},\boldsymbol{\tau}),\boldsymbol{\tau}\right)$. In actual applications, this will have to be transferred to the Eularian system's coordinates $\mathbf{r}$. Since $\mathbf{R}(\mathbf{r},\boldsymbol{\tau})$ is known, this should not present a problem.

The determination of the vector potential is done in two stages:

a) First, the vector potential is determined by Eq. (4.6) in a way similar to linear response theory. However the integral is not a convolution and cannot be performed in frequency space. One possible approximation is to use $\boldsymbol{N}(\mathbf{r},\boldsymbol{\tau}_0)$ instead of $\boldsymbol{N}(\mathbf{r},\boldsymbol{\tau}')$ in Eq. (4.6). $\boldsymbol{\tau}_0$ is a representative time (for example, $\boldsymbol{\tau}_0$ can be equal to $\boldsymbol{\tau}$). This may facilitate the computa-tion since the vector potential $\mathbf{A}_p$ will now be a convolution.

b) Once $\mathbf{A}_p$ is obtained, a transformation to the Eularian frame takes place, via Eq. (4.5). This transformation involves that Jacobians, which act as translation operators along the fluid parcel trajectory.

Memory is evident in Eq. (4.5), since the potential is sensitive to the past behavior of the velocity field. The final potentials do not depend on the Keldysh contour, as required. The fact that the scalar potential is zero is simply a specific choice of gauge. The longitudinal part of $\mathbf{a}_s = \mathbf{a}_{GIXC}/n$ can be converted to a scalar GIXC potential by an appropriate gauge transform.

It is interesting to note that Eq. (4.5) is consistent with the linear response theory of the homogeneous electron gas. We note that in linear response, we expand all quantities to first order. The first order change in density is $n_1$. The first order part of the velocity field $\mathbf{u}_1$ is the leading order (since we assume initial zero magnetic fields). This is also true of the vector potential. Therefore, $\Im \approx I + o(\mathbf{u}_1)$, $\mathbf{R}(\mathbf{r},t) \approx \mathbf{r} + o(\mathbf{u}_1)$, and Eq. (4.5) becomes similar to the time-domain form of the linear response result for a homogeneous electron gas[4]:

$$\begin{aligned}\mathbf{a}_{xc,1}(\mathbf{r},t) = &-\nabla\int_0^t \Phi_L(n_0,t-t')\nabla\cdot\mathbf{u}_1(\mathbf{r},t) + \\ &\nabla\times\int_0^t \Phi_T(n_0,t-t')\nabla\times\mathbf{u}_1(\mathbf{r},t),\end{aligned} \quad (4.8)$$

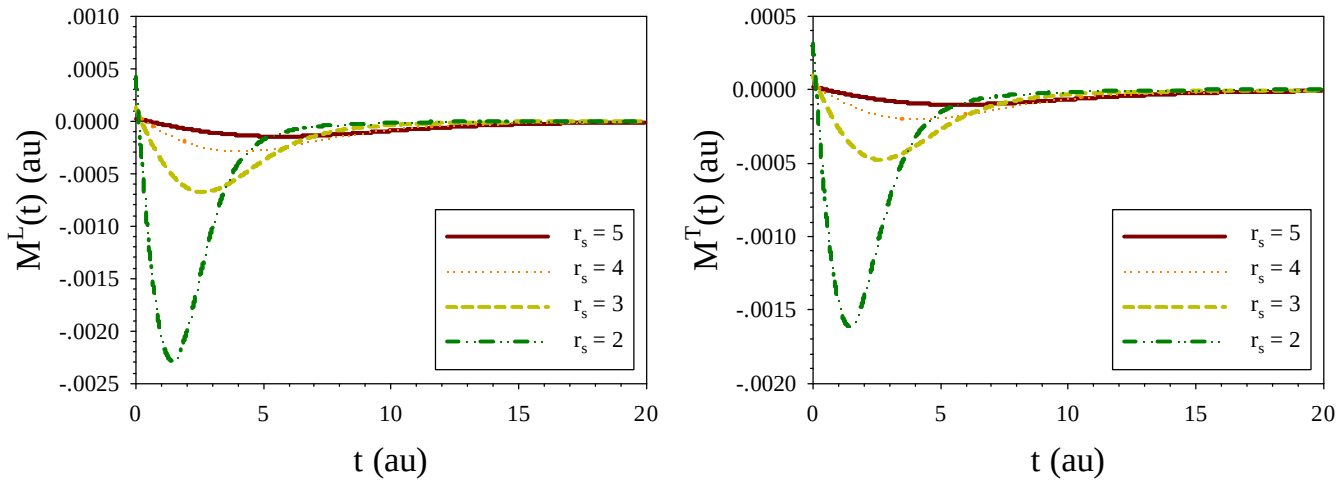

Figure 1: The longitudinal (left) and transverse (right) memory function kernels for various densities (values of $r_s$). Based on the LDA PW92 [36] functional and the Qian-Vignale $f_{xc}$ parameterization of [20].

## V. THE KERNEL FUNCTIONS

There are some exact results on the dynamical properties of the HEG in the LR regime. The relevance of these for TDDFT has been discussed in several references [4, 13, 18, 20, 37-39]. These known and extrapolated properties are encapsulated in the functions $f_{xc}^{L,T}(\omega)$, parameterization of which have been discussed [18, 20, 39]. The results we present are based on the $f_{xc}^{L,T}(\omega)$ of Qian et al[20].

In Appendix D we derive explicit expressions for the kernel functions, based on the known HEG response. The finite-memory kernel is derived:

$$M_{L,T}(N,t) = -\frac{2n^2}{\pi}\int_0^\infty \frac{f_i^{T,L}(N,\omega)}{\omega^2}\sin\omega\tau\, d\omega - \Phi_{L,T}^\infty(N) \quad (5.1)$$

In terms of which the kernels to be used in Eq. (4.6) are given by:

$$\Phi_L(N,t) = \Phi_L^\infty(N) + M_L(N,t) \quad (5.2)$$

$$\Phi_T(N,t) = \Phi_T^\infty(N) + \Phi_T^{ad}(N)t + M_T(N,t) \quad (5.3)$$

Where:

$$\begin{aligned}\Phi_{L,T}^\infty(N) &= -N^2 f_i'^{L,T}(N,0)\\ \Phi_T^{ad}(N) &= N^2 f_i^T(N,0)\end{aligned} \quad (5.4)$$

The memory functions $M_{L,T}(n)$ are shown in Figure 1, where the density is described by the Wigner Seitz parameter $r_s = \left(\frac{3}{4\pi n}\right)^{1/3}$. The adiabatic constant $\Phi_T^{ad}$ is shown in Figure 2.

Summarizing, we only need the time derivative of the Lagrangian vector potential $\mathbf{A}_p$ in Eq. (4.5), this is composed of 3 parts:

$$\dot{\mathbf{A}}_p(\mathbf{r},t) = \dot{\mathbf{A}}_p^{mem}(\mathbf{r},t) + \dot{\mathbf{A}}_p^\infty(\mathbf{r},t) + \dot{\mathbf{A}}_p^{ad}(\mathbf{r},t) \quad (5.5)$$

where the memory part is:

$$\begin{aligned}\mathbf{A}_p^{mem}(\mathbf{r},t) = &-\nabla\int_0^t dt' M_L\left(N(\mathbf{r},t'),t-t'\right)\nabla\cdot\mathbf{U}(\mathbf{r},t')\\ &+\nabla\times\int_0^t dt' M_T\left(N(\mathbf{r},t'),t-t'\right)\nabla\times\mathbf{U}(\mathbf{r},t'),\end{aligned} \quad (5.6)$$

Is evaluated by using a limited history of the fluid velocity and density because $M_{L,T}(n,t)$ is short ranged in $t$. The infinite response is given by:

$$\begin{aligned}\dot{\mathbf{A}}_p^\infty(\mathbf{r},t) = &-\nabla\left[\Phi_L^\infty(N(\mathbf{r},t))\nabla\cdot\mathbf{U}(\mathbf{r},t)\right]\\ &+\nabla\times\left[\Phi_T^\infty(N(\mathbf{r},t))\nabla\times\mathbf{U}(\mathbf{r},t)\right],\end{aligned} \quad (5.7)$$

Is evaluated with no need to refer to history. And the adiabatic part:

$$\dot{\mathbf{A}}_p^{ad}(\mathbf{r},t) = \nabla\times\int_0^t dt'\Phi_T^{ad}\left(N(\mathbf{r},t')\right)\nabla\times\mathbf{U}(\mathbf{r},t'), \quad (5.8)$$

which can be calculated "on the fly" with no need for memory, since this integral can be incremented at each time step.

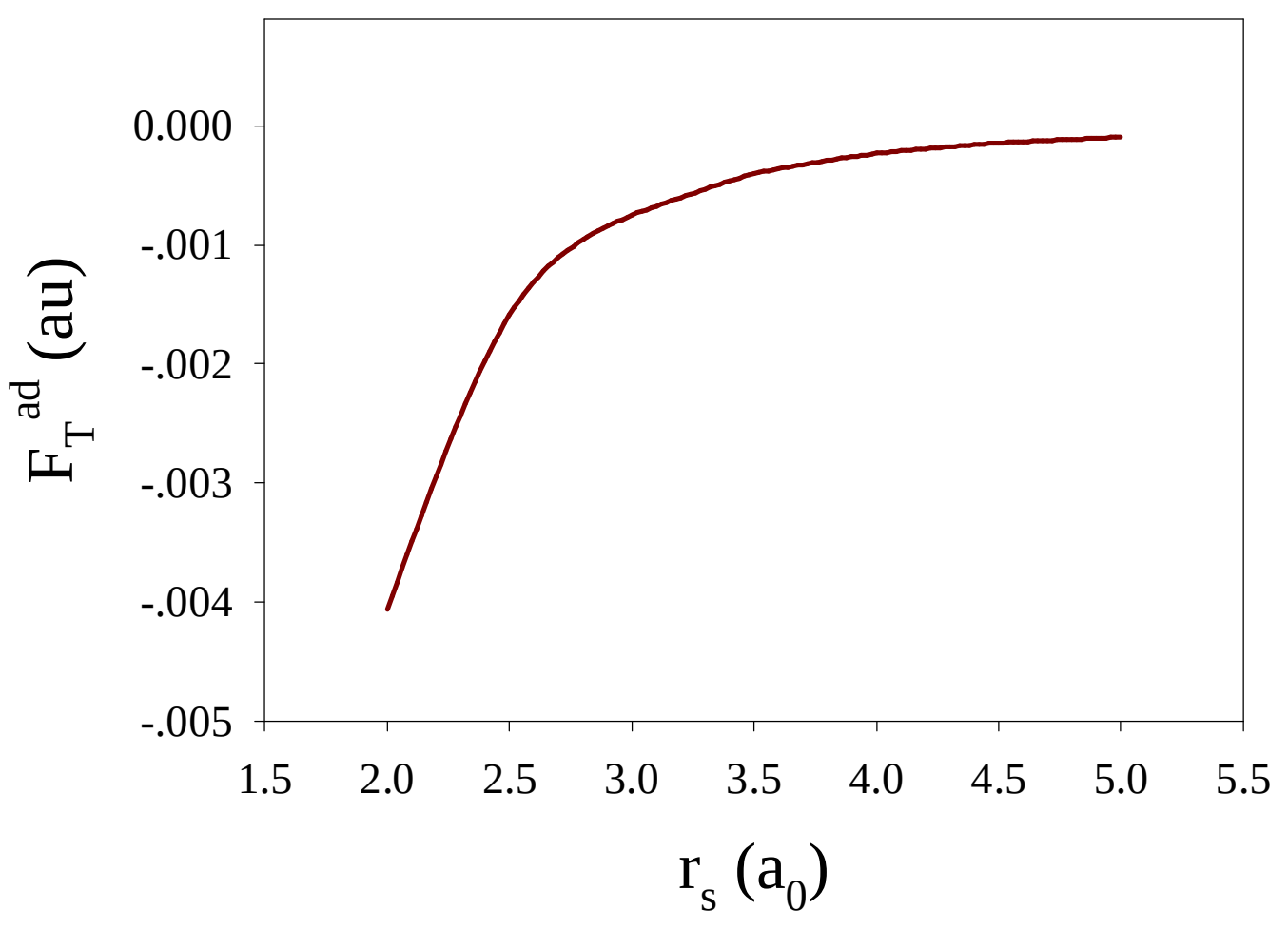

Figure 2: The constant $F_T^{ad}$ as a function of electron density (parameter $r_s$).

We have shown explicit kernels for the memory functional and discussed their application stressing the fact that all calculations are "doable", in the sense that they involve only a limited account of the history of the system.

## VI. SUMMARY AND DISCUSSION

In this paper we formulated an action approach to TDCDFT for electronic systems in an TD electromagnetic field, with the aim of constructing a memory action functional. Our formulation of the action is a generalization of the action devised for TDDFT[14]. We derived simple memory functionals that are robust due to their inherent Galilean Invariance and may be useful for taking into account memory effects in time-dependent calculations with strong fields. The use of a Lagrangian framework, as first suggested in Refs. [17] and [3] allows a full formulation of the memory effects in a Galilean Invariant way. The functionals are compatible with linear response properties, where the Lagrangian and Eularian frames are identical. Comparing with the theory of DBG (Ref. [3]), from which the present approach has been inspired, we find results are different in several aspects. While DBG derive their theory from an elegant application of Newton's third law, it is not clear if their potential can be derived from an action principle. The same question can be raised when comparing to the theory of Tokatly of Pankratov[11], which assume a that Lorenz force is a divergence of a stress tensor. In our treatment, we use a more general assumption and impose GI using a transformation from the Lagrangian to the Eularian frame. The importance of this method still needs to be examined by application to various benchmark systems. Furthermore, in the present approach the transverse part of the response is fully included.

In principle, the present approach can encompass more elaborate ansatz than the one introduced in Eq. (3.14). Future work will then address two issues related to this. A non-Newtonian-liquid approach should be attempted, where the derivatives of the Lagrangian velocity will be inserted in a non-linear way. Another issue is spatial locality. The current formalism is based on the local density approximation. Thus, there is no account of non-local spatial effects. While simply using the adiabatic GGA as a memory-less action functional in place of $S_A$ in Eq. (3.1) is a possibility, a more rigorous attempt, which combines non-locality in space and in time is still a challenge.

The incorporation of these equations in a real-time time-dependent scheme is more involved than the simple local-density approximation. However, the inclusion of memory effects may be the only way to improve the approximations we currently have for the dynamics of electrons in molecules.

While the present approach is far from addressing all the known properties of the true functional (for example the initial state dependence [40] is not addressed here), we believe it is a step forward, supplying a formalism that can be applied to general electronic structure systems.

**Acknowledgements** We gratefully acknowledge the support of the German Israel Foundation.

## APPENDIX A: CAUSALITY VIA THE KELDYSH CONTOUR

A special technique must be used to enforce causality in the action. Following van Leeuwen[14], the Keldysh time contour is used (although, in a different way). To explain the basic idea, let us define the Keldysh pseudo-time. Suppose the relevant interval for the physical time $\boldsymbol{t}$ is $[0,\boldsymbol{t}_f]$ ($\boldsymbol{t}_f$ can be infinity). We define a pseudo time $\tau \in [0,\tau_f]$ for some $\tau_f > 0$, and a parameterization $\boldsymbol{t}(\tau)$ which maps $[0,\tau_f]$ on $[0,\boldsymbol{t}_f]$ with $\boldsymbol{t}(0) = \boldsymbol{t}(\tau_f) = 0$. An important related concept is a "physical time" function: a function $f(\tau)$ ($f$ may depend on other variables as well) for which there exists a function $f_p(\boldsymbol{t})$ such that $f(\tau) = f_p(\boldsymbol{t}(\tau))$ is a "physical time function".

Peuckert[32], Rajagopal[33] and van-Leeuwen[14], describe an action, based on the Keldysh time contour, which by derivation produces causality-respecting potentials. Our approach to causality is also based on the Keldysh contour. Consider the following functional $A$ of a pseudo time-dependent function $n(\tau)$:

$$A[n] = \int_0^{\tau_f} \dot{\boldsymbol{t}}(\tau_2) d\tau_2 \times \int_0^{\tau_2} F\left(n(\tau_2), \boldsymbol{t}(\tau_2) - \boldsymbol{t}(\tau_1)\right) n(\tau_1) \dot{\boldsymbol{t}}(\tau_1) d\tau_1. \quad \text{(A.1)}$$

The functional derivative is:

$$\frac{dA[\nu]}{d\nu(\tau)} = \int_{\tau}^{\tau_f} \Phi\left(\nu(\tau_2), t(\tau_2) - t(\tau)\right)\dot{t}(\tau_2)d\tau_2 + \int_{0}^{\tau} \nu(\tau_1)\Phi_1\left(\nu(\tau), t(\tau) - t(\tau_1)\right)\dot{t}(\tau_1)d\tau_1 \qquad \text{(A.2)}$$

where $\Phi_1$ is the derivative of $\Phi$ by its first argument. By plugging in a "physical-time" function $\nu(\tau) = \nu_p(t(\tau))$ one obtains, after some manipulation:

$$v_p(t) \equiv \left.\frac{dA[\nu]}{d\nu(t)}\right|_p = \int_t^0 dt_2 \Phi\left(\nu(t_2), t_2 - t\right) + \int_0^t dt_1 \nu_p(t_1)\Phi_1\left(\nu(t), t - t_1\right) \qquad \text{(A.3)}$$

In deriving Eq. (A.3), we used the following fact: for any physical time function $f_p(t)$ and pseudo-times $\tau_1$ and $\tau_2$

$$t(\tau_1) = t(\tau_2) \Rightarrow \int_{\tau_1}^{\tau_2} f(t(\tau))\dot{t}(\tau)d\tau = 0. \qquad \text{(A.4)}$$

A crucial point: by choosing in Eq. (A.1), $\Phi(\nu, t)$ to be "causal" i.e. to be zero whenever $t < 0$, the functional derivative of (A.4) is reduced to a single term:

$$v_p(t) \equiv \left.\frac{dA[\nu]}{d\nu(t)}\right|_p = \int_0^t \nu_p(t_1)\Phi_1\left(\nu(t), t - t_1\right)dt_1. \qquad \text{(A.5)}$$

The potential $v_p(t)$ is now causal in the sense that the functional derivative $dv_p(t)/d\nu(t')$ is evidently zero when $t' > t$.

Another approach, more in the spirit of the Keldysh formalism[41] is to consider the time ordered functions on the contour, defining $\Phi^<(1,2)$ for $\tau_1 < \tau_2$ and $\Phi^>(1,2)$ for $\tau_1 > \tau_2$. Upon returning to physical quantities, only the combination $\Phi^R = \theta(t_2 - t_1)(\Phi^> - \Phi^<)$, called the retarded kernel, survives. Such an approach was taken by van-Leeuwen in his calculation of the response under TDDFT[14]. Thus, our formulation is akin to the retarded functional in the Keldysh theory.

## APPENDIX B: THE XC LORENTZ FORCE

In this appendix we show that the Galilean invariance of the XC action ensures zero net XC Lorentz force. The expression for the Lorenz force is given by Eq. (3.9), where the fields of Eq. (3.10) are used. Let us assume that Eq. (3.6) holds and prove that the net XC Lorentz force is zero. We mount a frame of reference slightly perturbed at some time $t > 0$ (see Eqs. (3.7)). The density and velocity fields appear slightly distorted in the perturbed frame (primed quantities):

$$\begin{aligned} \nu' &= \nu\left(\mathbf{r} + d\mathbf{x}(t'), t'\right), \\ \mathbf{u}' &= \mathbf{u}\left(\mathbf{r} + d\mathbf{x}(t'), t'\right) - \dot{t}^{-1}d\dot{\mathbf{x}}(t), \\ d\mathbf{x}(t') &= d\mathbf{x}\dot{t}^{-1}\delta(t - t') \end{aligned} \qquad \text{(B.1)}$$

From Galilean invariance, $\delta S_{xc} = S_{xc}[\nu', \mathbf{u}'] - S_{xc}[\nu, \mathbf{u}] = 0$ thus we have:

$$\int_C d\tau \int d^3r \left\{\frac{\delta S_{xc}}{\delta \nu}\nabla\nu \cdot d\mathbf{x} + \left(d\mathbf{x}\cdot\nabla\mathbf{u} - \dot{t}^{-1}d\dot{\mathbf{x}}\right)\cdot\frac{\delta S_{xc}}{\delta\mathbf{u}}\right\} = 0 \qquad \text{(B.2)}$$

Expressing this in indices, using the convention that repeated indices are summed, using Eq. (3.3) and multiplying by $-1$, for later convenience, the equation is transformed into:

$$\int \left\{-v\partial_j\nu - a_i\partial_j u_i - \dot{t}^{-1}\dot{a}_j\right\}dx_j d^3r = 0 \qquad \text{(B.3)}$$

Where $\mathbf{a} \equiv \mathbf{a}_{GIXC} \equiv \nu\tilde{\mathbf{a}}$ and $v \equiv v_{GIXC}$ for brevity. Integrating by parts the first term and using the fact that $dx_j$ is arbitrary, we have:

$$\int \left\{\nu\partial_j v - \nu\tilde{a}_i\partial_j u_i - \dot{t}^{-1}\left[\nu\dot{\tilde{a}}_j + \tilde{a}_j\dot{\nu}\right]\right\}d^3r = 0 \qquad \text{(B.4)}$$

Using the continuity equation, we have

$$\int \left\{\nu\left[\partial_j v - \dot{t}^{-1}\dot{\tilde{a}}_j\right] - \nu\tilde{a}_i\partial_j u_i + \tilde{a}_j\partial_i(\nu u_i)\right\}d^3r = 0 \qquad \text{(B.5)}$$

Since $\tilde{a}_i\partial_j u_i = \partial_j(\tilde{a}_i u_i) - u_i\partial_j\tilde{a}_i$, we have:

$$\int \left\{\nu\left[\partial_j(v - \tilde{a}_i u_i) - \dot{t}^{-1}\dot{\tilde{a}}_j\right] + \nu u_i\partial_j\tilde{a}_i + \tilde{a}_j\partial_i(\nu u_i)\right\}d^3r = 0 \qquad \text{(B.6)}$$

Integrating the last term by parts:

$$\int \left\{\nu\left[\partial_j(v - \tilde{a}_i u_i) - \dot{t}^{-1}\dot{\tilde{a}}_j\right] + \nu\left[u_i\partial_j\tilde{a}_i - u_i\partial_i\tilde{a}_j\right]\right\}d^3r = 0 \qquad \text{(B.7)}$$

Finally, we use the identity:

$$[\mathbf{u} \times (\nabla \times \tilde{\mathbf{a}})]_j = u_i\partial_j\tilde{a}_i - u_i\partial_i\tilde{a}_j \qquad \text{(B.8)}$$

And write:

$$\int \left\{\nu\left[\partial_j(v - \tilde{a}_i u_i) - \dot{t}^{-1}\dot{\tilde{a}} + [\mathbf{u} \times \nabla \times \tilde{\mathbf{a}}]_j\right]\right\}d^3r = 0 \qquad \text{(B.9)}$$

The integrand in the curly brackets is the average XC force per particle. Because $\nabla \times \tilde{\mathbf{a}}$ and $\nabla(v - \tilde{\mathbf{a}}\cdot\mathbf{u}) + \dot{\tilde{\mathbf{a}}}$ are gauge invariant this force too is gauge invariant. For the usual TDDFT (without vector potentials) this expression reduces to the expression of Vignale[16] $\int(-\nabla v)\nu d^3r = 0$.

## APPENDIX C: LAGRANGIAN FRAME

In this appendix we review several properties of the Lagrangian quantities. Consider the Lagrangian density

$N(\mathbf{r},t) = n(\mathbf{R}(\mathbf{r},t),t)$. Taking the time derivative, in obvious notation, we have:

$$\frac{\partial N(\mathbf{r},t)}{\partial t} = \frac{\partial n(\mathbf{R}(\mathbf{r},t),t)}{\partial t} + [\nabla n](\mathbf{R}(\mathbf{r},t),t)\cdot\mathbf{u}(\mathbf{R}(\mathbf{r},t),t)\dot{\tau}(t) \qquad \text{(C.1)}$$

Now, consider physical the density $n(\mathbf{r},t) = n_p(\mathbf{r},\tau(t))$, velocity field etc., then, with omission of the subscript $p$:

$$\frac{\partial N(\mathbf{r},t)}{\partial t} = \left[\frac{\partial n}{\partial t} + \mathbf{u}\cdot\nabla n\right](\mathbf{R}(\mathbf{r},t),t) \qquad \text{(C.2)}$$

Using the continuity equation(2.10), evaluated at $\mathbf{R}(\mathbf{r},t)$ we obtain:

$$\frac{\partial N(\mathbf{r},t)}{\partial t} + N(\mathbf{r},t)[\nabla\cdot\mathbf{u}](\mathbf{R}(\mathbf{r},t),t) = 0 \qquad \text{(C.3)}$$

Next, we consider the Jacobian

$$\Im_{ij}(\mathbf{r},t) = \partial R_i(\mathbf{r},t)/\partial r_j \qquad \text{(C.4)}$$

of the Eularian to Lagrangian transformation ($\mathbf{r} \to \mathbf{R}$). The Jacobian is needed because after formulation in the Lagrangian frame, we must transform back to the Eularian frame, where the other functionals (such as the adiabatic, Hartree and external functionals) are defined. Its determinant, which is used in various integrals, is also discussed.

In actual applications the Jacobian is readily available from the function $\mathbf{R}(\mathbf{r},t)$. In order to study its properties, it is instructive to obtain an equation of motion for it, which is obtained by taking the derivative of Eq. (C.4) with respect to $\mathbf{r}$:

$$\frac{\partial}{\partial t}\Im_{ij}(\mathbf{r},t) = \left.\frac{\partial u_i}{\partial r_k}\right|_{(\mathbf{R}(\mathbf{r},t),t)} \Im_{kj}(\mathbf{r},t)\dot{\tau}(t), \qquad \text{(C.5)}$$

Here we used Eq. (3.11) and the convention that repeated indices are summed over. Eq. (C.5) is a differential equation on $\Im$, which together with the initial condition that $\Im(\mathbf{r},0) = \mathbf{I}$, can be solved formally as:

$$\Im(\mathbf{r},t) = \wp\exp\left[\int_0^t [\nabla\mathbf{u}](\mathbf{R}(\mathbf{r},t'),t')\dot{\tau}(t')dt'\right] \qquad \text{(C.6)}$$

Where $\wp$ is a time ordering operator (earlier times appear to the right) and $[\nabla\mathbf{u}]_{ij} = \frac{\partial u_i}{\partial r_j}$.

Consider a small volume element $d^3r$. The number of particles in this element is $N(\mathbf{r},0)d^3r$. At time $t$ the element has moved to $\mathbf{R}(\mathbf{r},t)$, its shape and volume changed but the number of elements $N(\mathbf{r},t)|\Im(\mathbf{r},t)|^{-1}d^3r$ must still be the same, thus:

$$N(\mathbf{r},t)|\Im(\mathbf{r},t)|^{-1} = N(\mathbf{r},0) \qquad \text{(C.7)}$$

This equation is a useful way to compute the Jacobian determinant.

This result can be derived more rigorously from Eq. (C.6). Consider the determinant of a small time slice $\Delta t$ ($\Im$ is simply an ordered product of such slices, and the determinant of a product is the product of determinants). Because $|e^A| = e^{trA}$ for any operator, we have:

$$\left|e^{[\nabla\mathbf{u}](\mathbf{R}(\mathbf{r},t),t)\Delta t}\right| = e^{tr([\nabla\mathbf{u}](\mathbf{R}(\mathbf{r},t),t))\Delta t} \qquad \text{(C.8)}$$

However,

$$tr([\nabla\mathbf{u}](\mathbf{R}(\mathbf{r},t),t)) = (\nabla\cdot\mathbf{u})(\mathbf{R}(\mathbf{r},t),t) \qquad \text{(C.9)}$$

and using (C.3) we find:

$$tr([\nabla\mathbf{u}](\mathbf{R}(\mathbf{r},t),t)) = \frac{\partial}{\partial t}\ln N(\mathbf{r},t) \qquad \text{(C.10)}$$

And plugging into (C.8) gives:

$$\left|e^{[\nabla\mathbf{u}](\mathbf{R}(\mathbf{r},t),t)\Delta t}\right| = \frac{N(\mathbf{r},t)}{N(\mathbf{r},t-\Delta t)} \qquad \text{(C.11)}$$

Taking the product of all time slices we the Jacobian determinant equals

$$\frac{N(\mathbf{r},t)}{N(\mathbf{r},t-\Delta t)}\frac{N(\mathbf{r},t-\Delta t)}{N(\mathbf{r},t-2\Delta t)}\cdots\frac{N(\mathbf{r},\Delta t)}{N(\mathbf{r},0)} = \frac{N(\mathbf{r},t)}{N(\mathbf{r},0)}, \qquad \text{(C.12)}$$

confirming Eq. (C.7).

## APPENDIX D: THE RESPONSE KERNEL

In this appendix we discuss the construction of the memory kernel from the LR functions $f^{L,T}(n,\omega)$. To simplify the notation, we denote this function as $f(\omega)$, i.e. we drop the "L,T" super and subscripts and the explicit dependence on $n$.

Let us recall the definition of the function $f^L(\omega)$. It arises in the context of LR treatment of the homogeneous electron gas of density $n$. A weak perturbation by some external field, starting at $t = 0$ ensues a density response $n_1(\mathbf{r},t) = n(\mathbf{r},t) - n$. The time dependent density $n(\mathbf{r},t)$ is only slightly different from $n$ and $n_1(\mathbf{r},t)$ is proportional to the strength of the perturbation. The change in density $n_1(\mathbf{r},t)$ induces a change in the xc-potential $v_{1,xc}(\mathbf{r},t) = v_{xc}(\mathbf{r},t) - v_{xc}(\mathbf{r})$, depending linearly on $n_1$:

$$v_{1,xc}(\mathbf{r},t) = \int\int_0^t \Phi_{xc}(\mathbf{r}-\mathbf{r}',t-t')n_1(\mathbf{r}',t')dt'd^3r' \qquad \text{(D.1)}$$

Assuming further that a local density approximation is appropriate, i.e. $\Phi_{xc}(\mathbf{r}-\mathbf{r}',t-t') = \delta(\mathbf{r}-\mathbf{r}')\psi_{xc}(n(\mathbf{r}),t-t')$ we obtain:

$$v_{1,xc}(t) = \int_0^t y(t - t')n_1(t')dt', \quad \text{(D.2)}$$

where for notational clarity we drop the $n(\mathbf{r})$ dependence and the xc subscript. Also, for the HEG, $n$ is independent of $\mathbf{r}$. The function $y(t)$ is known to be composed of two parts. One is a function that has a Fourier transform $f(t)$ and the other is a delta function, associated with the infinite frequency response of the HEG:

$$y(t) = f(t) + f_\infty d(t) \quad \text{(D.3)}$$

Because of causality, let us assume explicitly that the function $f(t)$ is zero for $t < 0$ thus it is possible to extend the upper limit of integration to infinity:

$$v_{1,xc}(t) = \int_0^\infty f(t - t')n_1(t')dt' + f_\infty n_1(t). \quad \text{(D.4)}$$

Fourier transforming this convolution, we have:

$$v_{1,xc}(w) = f(w)n_1(w). \quad \text{(D.5)}$$

Where $n_1(w) = \int_{-\infty}^\infty n_1(t)e^{iwt}dt$ (with an analogous expression for $v_{1,xc}$) and

$$f(w) - f_\infty = \int_0^\infty f(t)e^{iwt}dt. \quad \text{(D.6)}$$

The fact that $f(w) - f_\infty$ is the Fourier transforms of a causal function poses a constraint on its analytical structure. Replacing $w$ by a complex frequency $z$, where $\mathrm{Im}(z) > 0$, on the right hand side of Eq. (D.6) yields a converging integral and thus constitutes an analytical continuation of $f(w)$ into the upper complex plane. From this fact, it is possible to derive Kramers-Kronig relations[42]:

$$f_r(w) - f_\infty = \frac{1}{p}P\int_{-\infty}^\infty \frac{f_i(w')}{w' - w}dw' \quad \text{(D.7)}$$

and

$$f_i(w) = -\frac{1}{p}P\int_{-\infty}^\infty \frac{f_r(w') - f_\infty}{w' - w}dw'. \quad \text{(D.8)}$$

Where $f$ is decomposed into its real and imaginary parts

$$f(w) = f_r(w) + if_i(w). \quad \text{(D.9)}$$

The function $f(w)$ contains also the adiabatic LDA response. This is encapsulated in the real zero frequency coefficient[18] $f(w = 0) = f_0(n) = \frac{d^2}{dn^2}(e_{xc}n)$, where $e_{xc}(n)$ is the xc-energy per particle for the HEG in its ground-state.

Because $f$ is real, the following is valid for $t > 0$:

$$\begin{aligned}\frac{p}{2}f(t) &= \int_{-\infty}^\infty f_i(w)\sin wt\,dw \\ &= \int_{-\infty}^\infty [f_r(w) - f_\infty]\cos wt\,dw.\end{aligned} \quad \text{(D.10)}$$

Using a gauge transformation in (D.2), we can transform $v_{1,xc}(t)$ into a vector potential:

$$\mathbf{a}_{1,xc}(t) = -\nabla\int_0^t dt'\int_0^{t'} y(t' - t'')n_1(t'')dt''. \quad \text{(D.11)}$$

Because $f(t)$ is causal, we have:

$$\mathbf{a}_{1,xc}(t) = -\nabla\int_0^t j(t - t'')n_1(t'')dt'', \quad \text{(D.12)}$$

where $j(t)$ is the causal function:

$$j(t) = \int_0^t f(t')dt' + f_\infty. \quad \text{(D.13)}$$

Using the second relation in (3.4), $\mathbf{a}_{1,GIXC} = n\mathbf{a}_{1,xc}$ we obtain an expression for the GIXC vector potential:

$$\mathbf{a}_{1,GIXC}(t) = -n\nabla\int_0^t j(t - t'')n_1(t'')dt'' \quad \text{(D.14)}$$

Let us now plug in the continuity equation (2.10), which in linear response regime is:

$$n_1(t) = -n\int_0^t \nabla\cdot\mathbf{u}_1(t')dt' \quad \text{(D.15)}$$

We have then:

$$\mathbf{a}_{1,GIXC}(t) = n^2\nabla\int_0^t j(t - t')dt'\int_0^{t'} \nabla\cdot\mathbf{u}_1(t'')dt'' \quad \text{(D.16)}$$

Integrating by parts, we have:

$$\mathbf{a}_{1,GIXC}(t) = \nabla\int_0^t x(t - t')\nabla\cdot\mathbf{u}_1(t')dt' \quad \text{(D.17)}$$

where,

$$x(t) = n^2\int_0^t j(t')dt', \quad \text{(D.18)}$$

and taking the time derivative, using Eq. (D.13):

$$\dot{x}(t) = n^2\left[\int_0^t dt' f(t') + f_\infty\right] \quad \text{(D.19)}$$

In other words we find that that in order for the kernel of Eq. (4.6) to be compatible with the LR properties of the HEG (assumed known as $f(t)$, or $f(w) - f_\infty$), the following must hold:

$$\ddot{x}(t) = n^2 f(t) \quad \text{(D.20)}$$

With the initial conditions:

$$x(0) = 0 \qquad \dot{x}(0) = n^2 f_\infty \quad \text{(D.21)}$$

Taking the Fourier transform of (D.20), using Eq. (D.6):

$$\int_0^\infty \ddot{x}(t)e^{i\omega t}dt = n^2[f(\omega) - f_\infty] \qquad \text{(D.22)}$$

And finally solving for $\ddot{x}(t)$ at $t > 0$ (see Eq. (D.10)):

$$\ddot{x}(t) = \frac{2n^2}{\pi}\int_0^\infty f_i(\omega)\sin\omega t\, d\omega \qquad \text{(D.23)}$$

The solution is therefore:

$$x(t) = a + bt - \frac{2n^2}{\pi}\int_0^\infty \frac{f_i(\omega)}{\omega^2}\sin\omega t\, d\omega \qquad \text{(D.24)}$$

and the constants $a$ and $b$ are selected to ensure $x(0) = 0$ and $\dot{x}(0) = n^2 f_\infty$, thus:

$$a = 0 \quad b = \frac{2n^2}{\pi}\int_0^\infty \frac{f_i(\omega)}{\omega}d\omega + n^2 f_\infty \qquad \text{(D.25)}$$

From the KK relations (D.7), $\frac{2}{\pi}\int_0^\infty f_i(\omega)/\omega\, d\omega = f_0 - f_\infty$, thus we have:

$$x(t) = n^2 f_0 t - \frac{2n^2}{\pi}\int_0^\infty \frac{f_i(\omega)}{\omega^2}\sin\omega t\, d\omega \qquad \text{(D.26)}$$

In order to check this result further, let us Fourier transform $x(t)$, obtaining $\tilde{x}(\omega) = -\left(n^2/\omega^2\right)f(\omega)$, and using it in $n\mathbf{a}_{1,xc} = \tilde{\mathbf{a}}_{1,GIXC}(\omega) = \nabla\left[\tilde{x}(\omega)\nabla\times\tilde{\mathbf{u}}_1(\omega)\right]$ - the Fourier version of Eq. (D.17), we obtain:

$$\mathbf{a}_{1,xc}(\omega) = -\frac{n}{\omega^2}\nabla\left[f(\omega)\nabla\times\mathbf{u}_1(\omega)\right] \qquad \text{(D.27)}$$

an expression that directly compares to results of ref. [4].

The term $n^2 f_0 t$ in (D.26) is linear in time and gives after gauging back to a potential, the adiabatic LDA potential, $v_{ALAD} = f_0 n_1$. This is because[18] $f_0(n) = \frac{d^2}{dn^2}(e_{xc}n)$. Thus it should not be part of the kernel $F$ in Eq. (4.6), since the adiabatic potential is obtained from the functional derivative of $S_A$ in Eq. (3.1). We find that the kernel in Eq. (4.6) must be given by:

$$F_L(t) = -\frac{2n^2}{\pi}\int_0^\infty \frac{f_i^L(\omega)}{\omega^2}\sin\omega t\, d\omega \qquad \text{(D.28)}$$

The function $F_L(t)$ does not decay to zero, instead we have:

$$\lim_{t\to\infty} F_L(t) = -n^2 f_i^{L\prime}(0) \equiv F_L^\infty \qquad \text{(D.29)}$$

This constant in time is treated separately, i.e. we define the finite memory functional, $F_L(t) = M_L(t) + F_L^\infty$ with:

$$M_L(t) = -\frac{2n^2}{\pi}\int_0^\infty \frac{f_i^L(\omega)}{\omega^2}\sin\omega t\, d\omega - F_L^\infty \qquad \text{(D.30)}$$

It is this kernel which is used in Eq. (4.6). The constant $F_L^\infty(n)$ is treated separately and because it is constant, its incorporation poses no problem.

A similar treatment, applies to $F_T$, only here we do not take out the ALDA part, thus:

$$F_T(t) = n^2 f_0^T t - \frac{2n^2}{\pi}\int_0^\infty \frac{f_i^T(\omega)}{\omega^2}\sin\omega t\, d\omega \qquad \text{(D.31)}$$

The behavior at large $t$ is thus:

$$F_T(t) \to n^2 f_0^T t - n^2 f_i^{T\prime}(0) \qquad \text{(D.32)}$$

So, writing: $F_T(t) = F_T^\infty + F_T^{ad}t + M_T(t)$, the finite memory functional is:

$$M_T(t) = -\frac{2n^2}{\pi}\int_0^\infty \frac{f_i^T(\omega)}{\omega^2}\sin\omega t\, d\omega - F_T^\infty \qquad \text{(D.33)}$$

where:

$$F_T^\infty = -n^2 f_i^{T\prime}(0) \qquad F_T^{ad} = n^2 f_i^T(0) \qquad \text{(D.34)}$$